\documentclass[num-refs]{wiley-article}

\usepackage{siunitx}
\usepackage{ragged2e}
\usepackage{setspace}

\title{On-Chip Neodymium-Doped Lithium Niobate Microdisk Laser with Self-Induced Pulsing}

\author[1]{Yuxuan He}
\author[1]{Jiangwei Wu}
\author[1]{Xiangmin Liu}
\author[1]{Feiyang Shen}
\author[2]{Feng Chen}
\author[2\authfn{2}]{Yuechen Jia}
\author[1,3]{Xianfeng Chen}
\author[1\authfn{2}]{Yuping Chen}

\affil[1]{School of Physics and Astronomy, State Key Laboratory of Advanced Optical Communication Systems and Networks, Shanghai Jiao Tong University, 800 Dongchuan Road, Shanghai 200240, China}
\affil[2]{School of Physics, State Key Laboratory of Crystal Materials, Shandong University, Jinan 250100, China}
\affil[3]{Collaborative Innovation Center of Light Manipulations and Applications, Shandong Normal University, Jinan 250358, China}

\corraddress{Yuechen Jia, School of Physics, State Key Laboratory of Crystal Materials, Shandong University, Jinan 250100, China

Yuping Chen, School of Physics and Astronomy, State Key Laboratory of Advanced Optical Communication Systems and Networks, Shanghai Jiao Tong University, 800 Dongchuan Road, Shanghai 200240, China
}

\corremail{yuechen.jia@sdu.edu.cn, ypchen@sjtu.edu.cn}

\presentadd[\authfn{2}]{School of Physics and Astronomy, State Key Laboratory of Advanced Optical Communication Systems and Networks, Shanghai Jiao Tong University, 800 Dongchuan Road, Shanghai 200240, China}

\runningauthor {Yuxuan He et al.}

\begin{document}

\begin{frontmatter}
\maketitle

\begin{abstract}
\justifying 
\begin{spacing}{1.11}

Rare-earth–doped materials constitute the foundation of conventional solid-state lasers, but their bulk-crystal form is inherently incompatible with photonic integration, making it challenging to realize compact, high performance nanoscale laser sources. Lithium niobate on insulator (LNOI), with its exceptional electro-optic and nonlinear optical properties, has emerged as one of the most promising platforms for integrated photonics. Combining Nd$^{3+}$ doping with LNOI offers the unique possibility of uniting the efficient gain provided by Nd$^{3+}$ ions with the excellent characteristics of LNOI. However, on-chip laser emission from Nd:LNOI has not been demonstrated previously.
In this work, we report the first realization of an integrated Nd:LNOI microdisk laser, demonstrating lasing at 1094.17 nm under 785.10 nm pumping with a low threshold of 146 $\mu$W and a slope efficiency of $1.962\times10^{-5}$. Beyond continuous-wave operation, we further observe self-induced laser pulsing on the hundred-microsecond scale, with a laser-pulse duration down to 500 $\mu$s and an oscillation period of 6.45 ms, arising from nonlinear thermo-optic–photorefractive dynamics. We demonstrate stable continuous wave lasing and self-induced pulsed emission within a monolithically integrated Nd:LNOI cavity. Our results expand the operational degrees of freedom for LNOI-based lasers and open a new direction toward deeply integrated gain with intrinsic nonlinear dynamical processes.
\end{spacing}
\justifying 

\end{abstract}
\end{frontmatter}

\section{Introduction}
Rare-earth–ion-doped solid-state lasers have long occupied a central position in modern photonics owing to their high efficiency, excellent spectral purity, and long energy storage lifetimes\cite{li2023optically, fang2024high, luo2023advances, liu2021chip, liu2025ultralow, zhang2021integrated}. Among them, neodymium ions (Nd$^{3+}$) are particularly important because of their unique four-level transition scheme and their rich emission bands spanning the near-infrared region (900–1400 nm), which underpin numerous high-power and narrow-linewidth laser systems. The 4F$_{3/2}$ $\rightarrow$ 4I$_{11/2}$ transition of Nd$^{3+}$ exhibits a large stimulated emission cross section, making about 1.1 $\mu$m the dominant lasing wavelength. Meanwhile, the relatively long upper-level lifetime of the 4F$_{3/2}$ state (typically hundreds of microseconds) allows efficient population inversion under moderate pump intensities\cite{jia2022integrated, shen2024mode}. These attributes have made Nd-doped media become highly favorable gain materials for both bulk and microcavity solid-state lasers.

Over the past decades, a variety of host crystals capable of incorporating Nd$^{3+}$ ions—such as Y$_{3}$Al$_{5}$O$_{12}$ (YAG), YVO$_{4}$, YAlO$_{3}$ (YAP), and GdVO$_{4}$—have been extensively developed to realize efficient solid-state laser emission\cite{li2023optically, de2022nd, yu2025orthogonally, song2021excellent}. These hosts offer wide optical transparency windows, high thermal conductivity, and mature crystal growth techniques\cite{guo2021growth, turri2022mid, si2024multiphonon} and thus have become the primary gain media for commercial Nd-doped lasers. However, the intrinsic physical and structural properties of these bulk materials fundamentally limit their compatibility with modern photonic integration. Conventional Nd-doped crystals are typically grown by the Czochralski methods\cite{saleh2020improved}, yielding centimeter-scale single crystals whose fabrication relies on mechanical cutting and polishing. Due to their high melting point, strong ionic bonding, and rigid lattice structures, these materials are difficult to transform into high-quality thin films by low-temperature epitaxy, pulsed laser deposition (PLD), or ion-slicing techniques\cite{saleh2020improved, zajic2023nd}. Consequently, they are unsuitable for nanoscale or heterogeneous photonic integration. In addition, common hosts such as YAG and YVO$_{4}$ exhibit weak second-order nonlinear coefficients or lack electro-optic tunability, confining their functionality to optical gain alone and preventing the realization of fully integrated active–passive photonic platforms.

In contrast, lithium niobate (LiNbO$_{3}$) is a multifunctional ferroelectric crystal that combines excellent electro-optic and nonlinear optical properties, and has long been utilized in frequency conversion, electro-optic modulation, and quantum photonic devices\cite{zhu2021integrated}. Recent advances in ion-slicing and direct wafer bonding technologies have led to the emergence of the lithium niobate on insulator (LNOI) platform, which dramatically expands the application scope of LiNbO$_{3}$. Single-crystalline thin films with submicrometer thicknesses now exhibit propagation losses below 0.1 dB/cm and strong optical confinement\cite{zhu2024twenty}. These films retain the superior nonlinear and electro-optic properties of bulk LiNbO$_{3}$ while being fully compatible with planar nanophotonic fabrication, thus establishing a solid foundation for large-scale integrated photonics.

To further endow the LNOI platform with active optical gain functionality, researchers have sought to incorporate Nd$^{3+}$ ions into LNOI. However, achieving effective Nd$^{3+}$ doping in LNOI presents significant technical challenges. Traditional diffusion-based rare-earth doping requires temperatures exceeding 1000 °C—far beyond the thermal stability limit of LNOI—while ion implantation followed by high-temperature annealing is likewise constrained by thermal damage, leading to limited doping depth, nonuniform spatial distribution, and lattice defects.

A recent study has successfully demonstrated an alternative route: performing high-temperature Nd$^{3+}$ diffusion into bulk LiNbO$_{3}$ prior to film fabrication, followed by the realization of thin-film Nd:LNOI\cite{ruter2021optical}. This approach enables uniform Nd$^{3+}$ incorporation while preserving the crystalline quality of the LNOI layer. Compared with other hosts, Nd:LNOI not only inherits the distinctive amplification and emission characteristics of Nd$^{3+}$ but also fully benefits from the rich optical functionalities of LNOI itself. More importantly, it exhibits functional compatibility with the rapidly evolving LNOI integrated photonic platform. These advantages make Nd:LNOI an ideal material system for realizing compact on-chip lasers and exploring the interplay between Nd$^{3+}$ optical gain and the intrinsic nonlinear dynamics of LiNbO$_{3}$.

Despite extensive research on rare-earth-doped thin-film lithium niobate lasers, lasing emission from Nd$^{3+}$ on the LNOI platform has not yet been reported. In this work, we demonstrate, for the first time, an integrated Nd:LNOI laser operating at 1094.17 nm under 785.10 nm optical pumping. The device can generate both continuous-wave and self-induced pulsed laser outputs. This achievement extends the emission wavelength of active LNOI photonic platforms toward the 1.1 $\mu$m near-infrared region and opens new opportunities for realizing multifunctional on-chip lasers governed by nonlinear thermo-optic–photorefractive dynamics.

\section{CONTINUOUS WAVE LASER OUTPUT}

Neodymium ions (Nd$^{3+}$) exhibit a well-defined four-level energy structure that enables efficient laser emission in various host materials. The ground state 4I$_{9/2}$ is comprised of multiplets which can be optically pumped to 4F$_{5/2}$ and 2H$_{9/2}$ under pumping near 790 nm, corresponding to large absorption cross section. The excited states rapidly relax nonradiatively to the metastable energy level of 4F$_{3/2}$, which serves as the upper laser level with a relatively long lifetime. The laser transition occurs from 4F$_{3/2}$ $\rightarrow$ 4I$_{11/2}$, giving rise to emission around 1100 nm as shown in figure 1(a), depending on the host crystal field. Subsequently, the population relaxes quickly from 4I$_{11/2}$ to the ground state through multi-phonon processes, completing the laser cycle.

We fabricated a microdisk cavity with a diameter of approximately 104 $\mu$m on a 600 nm-thick Nd:LNOI using the femtosecond laser ablation–assisted chemical mechanical polishing (FLA–CMP) technique\cite{wu2018lithium}. Microdisks fabricated using this method exhibit very strong cavity enhancement, which is favorable for various on-chip nonlinear effects\cite{fu2024soliton} and for laser generation. The detailed fabrication process is provided in the Supplementary Information (SI), and the SEM image of the fabricated microdisk is shown in Fig. 1(b). The transmission spectrum of the microdisk, shown in Fig. 1(c) and 1(d), reveals a Q factor of $6.48\times10^5$ obtained through lorentz fitting. This relatively high Q factor facilitates low-threshold lasing emission from the Nd-doped microdisk cavity.

\begin{figure}[bt]
\centering
\includegraphics[width=\textwidth]{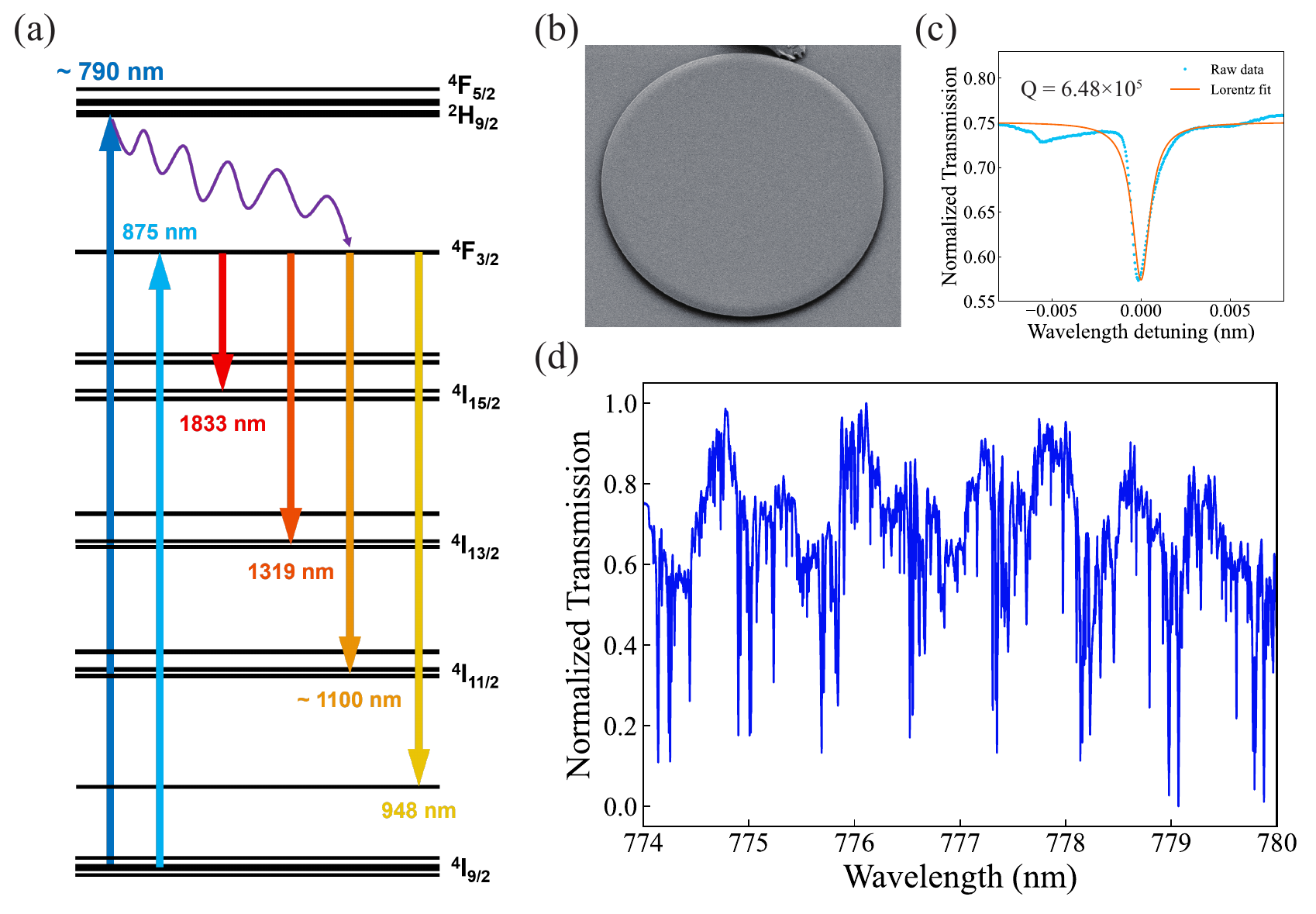}
\caption{(a) Energy level illustration of Nd$^{3+}$. (b) SEM image of Nd:LNOI microdisk. (c) Transmission spectrum of Nd:LNOI microdisk from 774 nm to 780 nm. (d) Q factor of about $6.48\times10^5$ extracted from the transmission spectrum.}
\end{figure}

In experiment, under 785.10 nm pumping, a single-peak laser output at 1094.17 nm was observed. Meanwhile, the spectral distribution between 1100 and 1125 nm exhibited a low intensity broadband emission corresponding to the Nd$^{3+}$ gain spectrum in this wavelength range. However, due to the relatively weaker gain intensity in this wavelength range compared with the lasing band, the emission in this region appears as fluorescence rather than stimulated emission.

\begin{figure}[bt]
\centering
\includegraphics[width=\textwidth]{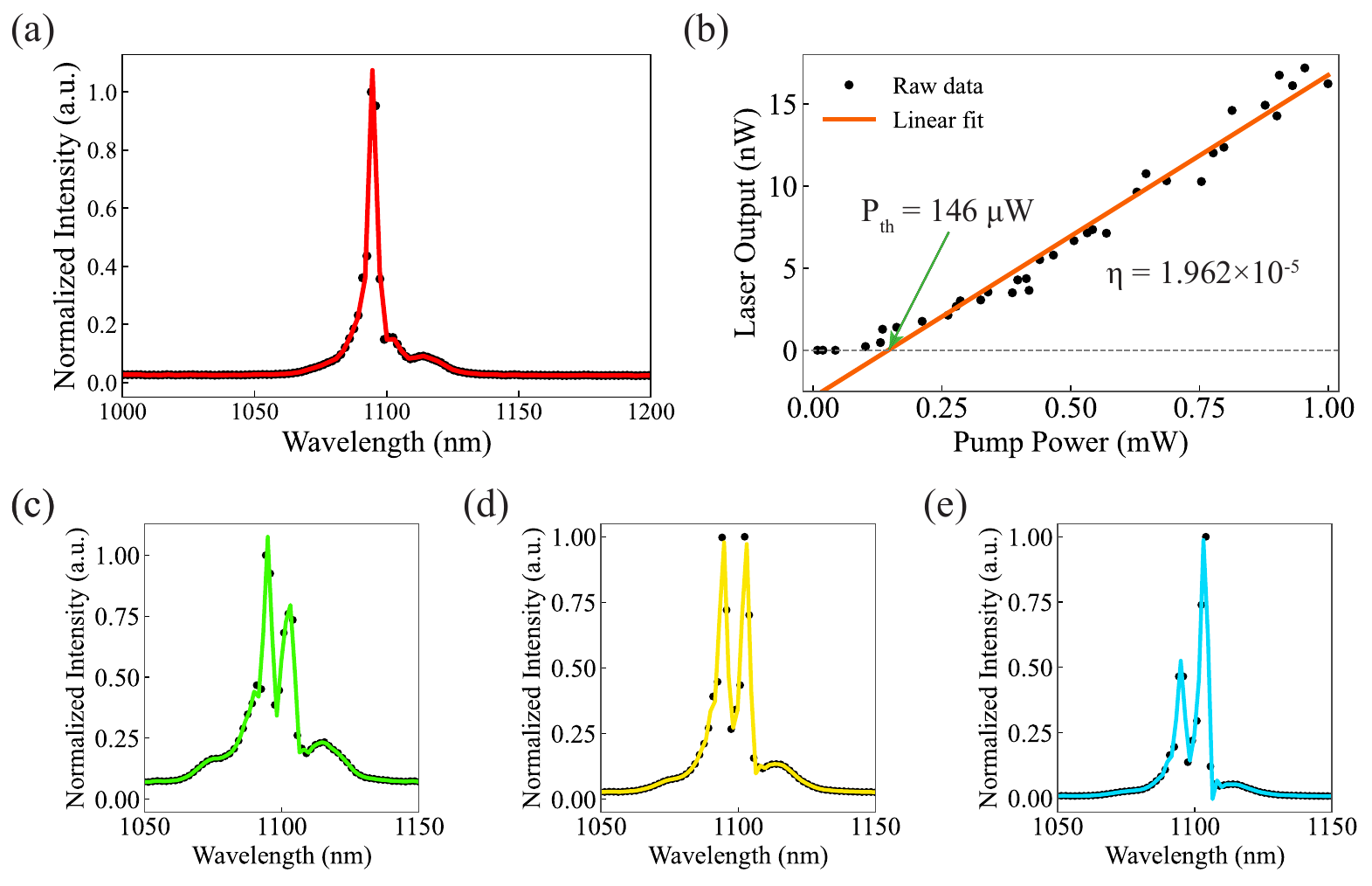}
\caption{(a) Single-peak laser obtained from Nd:LNOI microdisk with pump wavelength of 785.10 nm. (b) Linear power relationship with a slope efficiency of $1.962\times10^{-5}$ and laser threshold of about 146 $\mu$W. Multi-peaks laser shapes of (c), (d) and (e) obtained from Nd:LNOI microdisk demonstrating relative-intensity adjustment of the two lasing peaks. The black points correspond to the raw data, and the colored curves represent the fitted results.}
\end{figure}

The Nd:LNOI laser exhibits a threshold of 146 $\mu$W and a slope efficiency of $1.962\times10^{-5}$, which is comparable to that of most on-chip Er-doped lasers. When the pump wavelength is further tuned to 782.73 nm, the laser output evolves into a distinct dual-peak spectrum: in addition to the original 1094.17 nm peak, a second lasing peak emerges at 1102.39 nm. By finely tuning the pump wavelength within a very narrow range, the relative-intensity of the two lasing peaks can be adjusted, as shown in Fig. 2(c)–(e). This behavior arises from mode competition among higher-order modes within the microdisk cavity, a consequence of its small sidewall angle. A detailed discussion is provided in the SI.

\section{SELF-INDUCED PULSING}
In addition to the multimode, low-threshold continuous-wave (CW) lasing, we also achieved on-chip pulsed lasing from the Nd:LNOI platform.
Pulsed lasers are of great importance in optical communications, time-resolved spectroscopy, LiDAR, and nonlinear microscopy, where precise temporal control and high peak power are required\cite{song2025stable, nie2025soliton, he2023massively}.

In conventional laser systems, pulsed operation can be realized through various techniques, such as active Q-switching driven by electro-optic or acousto-optic modulation, or passive Q-switching and mode-locking using saturable absorbers\cite{rosol2024short, najm2023generation, wang2025wgm}.
Beyond these approaches, self-induced pulsing refers to a regime in which a laser spontaneously oscillates between gain and loss without any external modulation, generating temporally modulated pulses purely through intrinsic system dynamics\cite{wu2025manipulation}.
Such self-pulsing behavior typically arises from internal nonlinearities that dynamically couple the intracavity optical field and the resonator environment, including thermo-optic, Kerr, or photorefractive effects\cite{sun2017nonlinear, wang2023nanosecond}.

Our Nd:LNOI microdisk resonators exhibit strong thermo-optic and photorefractive nonlinearities.
The thermo-optic effect is a relatively fast process ($\sim$10 $\mu$s), where the rising temperature increases the effective refractive index, leading to a redshift of the cavity resonance.
In contrast, the photorefractive effect is a much slower process($\sim$1 ms). The migration of photo-induced charge carriers forms a space-charge field which decreases the refractive index through the electro-optic response of LN and thereby causes a blueshift of the resonance.
When the pump power exceeds a certain threshold, the interplay between these two competing nonlinearities periodically modulates the cavity resonance and the intracavity gain, giving rise to self-sustained pulsed laser emission without any external modulation.

Driven by this nonlinear thermo-optic–photorefractive dynamics, we observed this self-induced pulsed laser in the Nd:LNOI microdisk with a pulse duration on the order of hundreds of microseconds.
This observation reveals a unique dynamic coupling between optical gain and intrinsic nonlinearities in the Nd:LNOI platform, and provides a fundamental basis for further performance optimization of on-chip pulsed lasers based on Nd:LNOI.

\begin{figure}[bt]
\centering
\includegraphics[width=\textwidth]{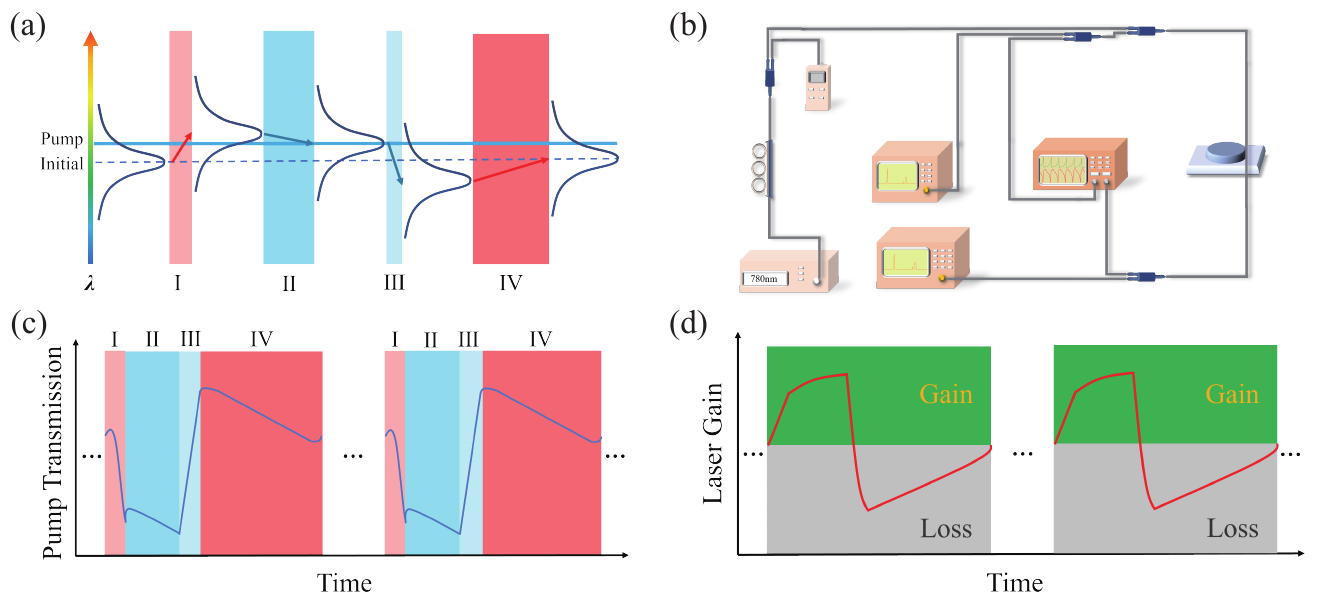}
\caption{(a) Schematic of the cavity resonance shift induced by the thermo-optic-photorefractive mechanism varied with time. (b) Setup of nonlinear pulsed laser oscillation experiment. (c) Periodical waveform of cavity transmission with the four stages denoted in Fig. 3(a). (d) Periodical modulation of laser gain with alternate gain and loss.}
\end{figure}

The nonlinear thermo-optic–photorefractive dynamics in the Nd:LNOI microdisk proceeds through four characteristic phases, as depicted in Fig. 3(a). When the excitation laser is launched on the red-detuned side of a cold-cavity resonance, the resonance experiences a rapid thermally driven red shift because the thermo-optic (TO) response is considerably faster than the photorefractive (PR) effect. After this thermal shift, the input wavelength effectively becomes blue-detuned with respect to the new, red-shifted resonance.
In stage II, the gradually accumulated space-charge distribution gives rise to an electro-optic change of the refractive index, which moves the cavity mode toward shorter wavelengths and slowly compensates the initial thermal red shift. This compensation continues until the wavelength mismatch between the cavity mode and the pump laser vanishes, at which point the coupling efficiency reaches its maximum. As the PR effect further strengthens, stage III begins and the resonance continues to move to the shorter wavelengths, reducing the intracavity optical power. With the optical heating effect weakened, the TO contribution decays rapidly, and the detuning between pump and cavity mode reaches its largest value.
In Stage IV, the stored space-charge field starts to relax because of the low optical power. The resonance then gradually shifts toward longer wavelengths, eventually recovering the initial position corresponding to the beginning of stage I. These four stages complete a full oscillatory cycle. A more comprehensive theoretical description is provided in the SI.

Because both the TO and PR effects are governed by the intracavity optical power, the onset of nonlinear oscillation requires the pump power to exceed a certain threshold. Once the threshold is achieved, the cavity resonance begins to undergo periodic ocsillation, leading to a periodic modulation of the transmission of pump power, as illustrated in Fig. 3(c). This indicates that the pump power coupled into the microdisk varies periodically in time.
When the pump wavelength lies within the absorption band of Nd$^{3+}$ ions, the gain within the corresponding emission band is likewise periodically modulated. The temporal oscillation of the gain follows the same period as the oscillation observed in the pump transmission, as shown in Fig. 3(d). During the intervals marked in green in Fig. 3(d), the instantaneous gain exceeds the lasing threshold and laser emission occurs; conversely, in the gray regions, the gain falls below threshold and lasing is suppressed. This periodic switching of the net gain results in the generation of pulsed laser output.

In experiment, a tunable 780 nm laser was employed as the pump source.
The pump was delivered to the Nd:LNOI microdisk cavity after passing through a polarization controller and a power meter for polarization adjustment and power monitoring.
To accurately observe the synchronous oscillation between the signal and pump light, the backward-propagating signal was collected and directed through two cascaded wavelength division multiplexers (WDMs) to ensure efficient spectral separation of the signal and residual pump components as shown in figure. 3(b).
The filtered signal light, together with the forward-propagating pump light, was then monitored using a high-speed oscilloscope.
Since the intensity of the pump light was much stronger than that of the signal, only one WDM was inserted in the forward path for pump filtering.
Additionally, spectrometers were connected to both the reflection and transmission ports to verify the spectral purity and confirm the effectiveness of the filtering setup.

\begin{figure}[bt]
\centering
\includegraphics[width=\textwidth]{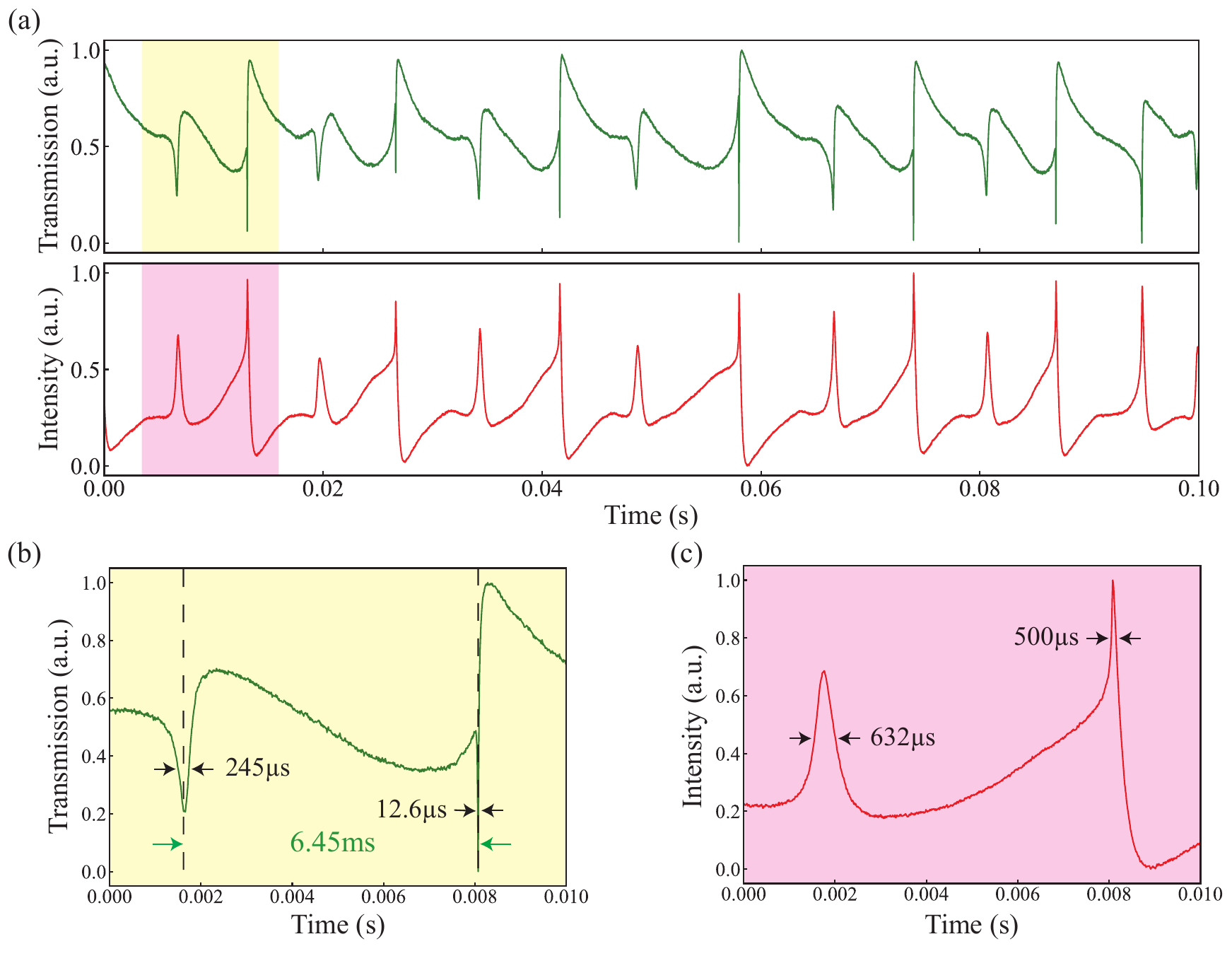}
\caption{(a) Synchronous transmission spectrum of the periodical pulsed pump and laser. The green curve represents the pump transmission spectrum and the red curve represents the laser emission spectrum. (b) Zoom in area of the transmission spectrum. The oscillation period is about 6.45 ms. The duration of pump can reach a low level of 12.6$\mu$s. (c) Zoom in area of the laser emission spectrum. The duration of the laser can reach a low level of 500$\mu$s.}
\end{figure}

When the pump power was increased to the milliwatt level, self-induced pulsed laser emission began to emerge. A representative measurement is shown in Figure 4(a). The upper green trace corresponds to the periodic oscillation of the pump transmission, while the lower red trace shows the output of the pulsed laser. The two waveforms exhibit clear temporal synchronization, and each oscillation cycle consists of two adjacent peaks (or dips). This behavior is likely associated with the presence of two cavity resonances located in close proximity to the pump wavelength. The nonlinear thermo-optic–photorefractive dynamics simultaneously drives the wavelength drift of both modes, giving rise to a dual-mode oscillation pattern. The existence of high-order transverse modes or the mode splitting due to the surface nonuniformities of the microdisk may be the reasons.
Fig. 4(b) and 4(c) present enlarged views of the oscillation features. The temporal separation between two adjacent pulses is measured to be 6.45 ms. The minimum dip width of the pump transmission reaches 12.6 $\mu$s, corresponding to a duty ratio of $1.95\times10^{-3}$. The emitted laser pulses exhibit a broadened linewidth of approximately 500 $\mu$s, with a duty ratio of only $7.75\times10^{-2}$ which represents more than one order-of-magnitude improvement compared to our earlier results\cite{wu2025manipulation}. Such low-duty-ratio operation provides a promising pathway toward high-peak-power, low-average-power pulsed laser systems with precise temporal control.

\begin{figure}[bt]
\centering
\includegraphics[width=\textwidth]{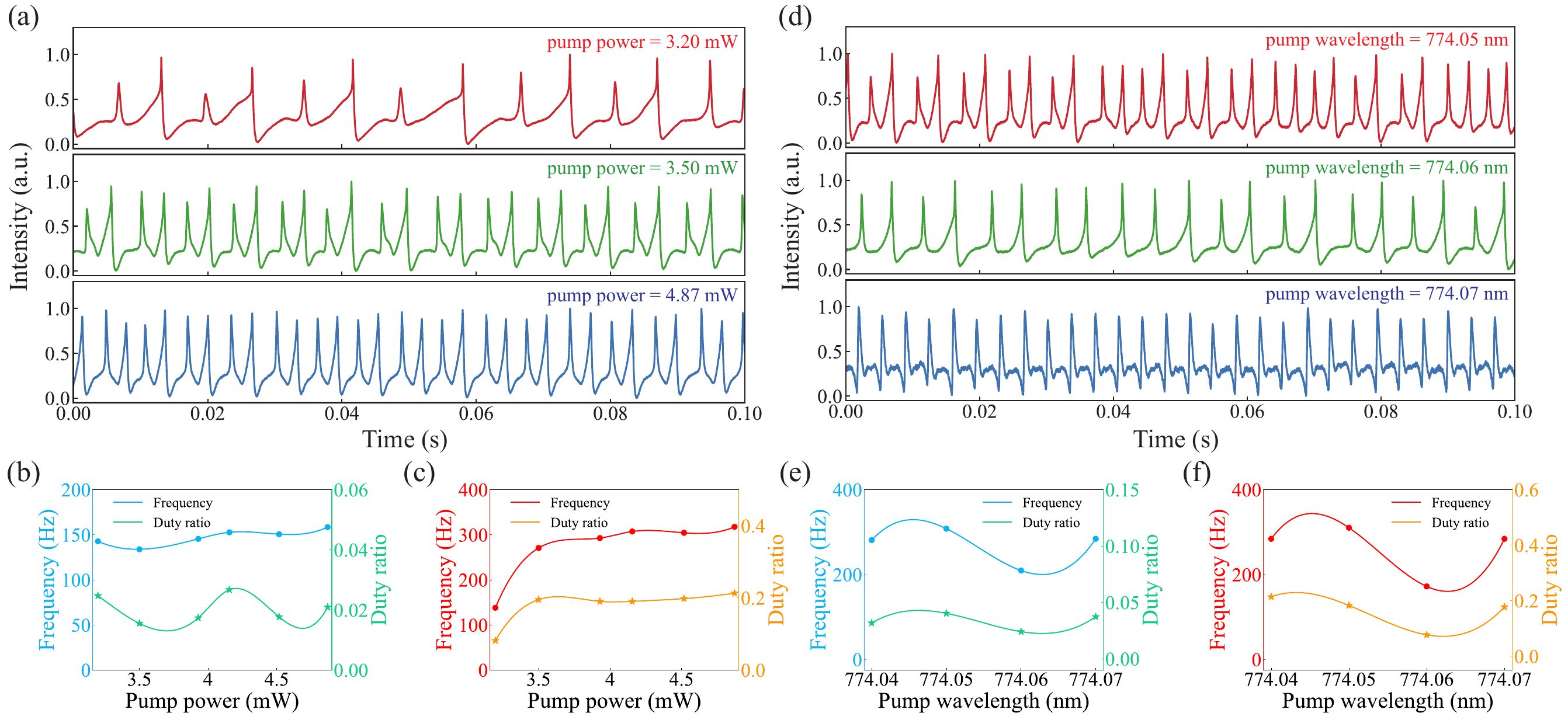}
\caption{Periodical pulsed laser emission with (a) the pump power from 3.198 mW to 4.870 mW and (d) the pump wavelength from 774.04 nm to 774.07 nm. (b), (c), (e) and (f) demonstrate the nonlinear oscillation properties. (b) and (e) show the variation of oscillation frequency and duty ratio with the pump power. (c) and (f) show the variation of oscillation frequency and duty ratio with the pump wavelength.}
\end{figure}

In addition, we investigated how the pump power and pump wavelength influence the pulsed laser dynamics. Figures 5(a) and 5(d) show representative temporal waveforms of the pulsed emission under different pump powers and wavelengths, while Figures 5(b), 5(c), 5(e), and 5(f) summarize the corresponding variations in pulse repetition frequency and duty ratio. As illustrated in Figure 5(a), when the pump power increases from 3.20 mW to 3.50 mW, the oscillation frequency nearly doubles. Beyond this range, both the pulse frequency and duty ratio become largely insensitive to further increases in pump power. From Figure 5(d), we observe that both the frequency and duty ratio of the pulsed laser show a trend of first decreasing and then increasing as the pump wavelength increases. This behavior may originate from the intricate nonlinear thermo-optic–photorefractive dynamics in the microdisk cavity, and warrants further investigation.

\section{DISCUSSION}
In summary, we have demonstrated the first on-chip Nd-doped LNOI laser operating at 1094.17 nm, featuring a relatively low lasing threshold of 146 $\mu$W and a slope efficiency of $1.962\times10^{-5}$. These results confirm the potential of Nd:LNOI as a new gain medium in integrated photonics and provide more possibilities for the integrated LNOI photonics platform. Furthermore, by investigating the nonlinear thermo-optic-photorefractive oscillation dynamics, we observe stable self-induced pulsed lasing with durations on the order of hundreds of microseconds. This nonlinear pulsing behavior introduces a new dynamical dimension to on-chip laser study in LNOI.
Overall, our results indicate that Nd:LNOI serves not only as an efficient on-chip gain platform, but also as a promising medium for exploring integrated nonlinear optics and complex dynamical phenomenon. Through combining with new techniques such as sectional doping or others\cite{chen2022photonic}, the demonstrated system can open up opportunities for chip-scale pulsed light sources, photonic neuromorphic elements, self-modulated oscillators, and integrated active–nonlinear photonic circuits.

\section*{acknowledgements}
This work was supported by the National Natural Science Foundation of China (12134009 and 12474335). The authors thank Prof. Ya Cheng and Prof. Jintian Lin for their help of chemo-mechanical polishing. The authors thank Prof. Xiaoqin Shen at ShanghaiTech University for his help of experiment and the Center for Advanced Electronic Materials and Devices of Shanghai Jiao Tong University for its support in device fabrication.

\section*{conflict of interest}
The authors declare no conflict of interest.

\section*{data availability statement}
The data that support the findings of this study are available from the corresponding author upon reasonable request.

\section*{Supporting Information}

The Supporting Information is available free of charge and contains four parts as shown below.

Fabrication process of Nd:LNOI microresonators, multi-mode characteristics of Nd:LNOI microdisk resonator, second harmonic generation of Nd:LNOI microresonators, dynamic equations for nonlinear oscillation of thermo-optic-photorefractive effect (PDF)

\printendnotes
\bibliography{sample}

@article{fang2024high,
  title={High-power Pr\^{} 3+: YLF continuous wave lasers at 691.7 nm, 701.4 nm, 705.0 nm, and 708.7 nm},
  author={Fang, Run and Dai, Rongbin and Xu, Huiying and Cai, Zhiping},
  journal={Chinese Optics Letters},
  volume={22},
  number={8},
  pages={081404},
  year={2024}
}

@article{shen2024mode,
  title={Mode-locked fiber lasers at 1064 and 910 nm wavelengths using Nd\^{} 3+-doped silica fiber},
  author={Shen, He'nan and Zhao, Xilong and Yu, Fei and Wang, Yazhou and Wang, Yafei and Sun, Yan and Wang, Shikai and Chen, Yinggang and Jia, Zhongqing and Zhai, Ruizhan and others},
  journal={Chinese Optics Letters},
  volume={22},
  number={9},
  pages={091402},
  year={2024}
}

@article{luo2023advances,
  title={Advances in lithium niobate thin-film lasers and amplifiers: a review},
  author={Luo, Qiang and Bo, Fang and Kong, Yongfa and Zhang, Guoquan and Xu, Jingjun},
  journal={Advanced Photonics},
  volume={5},
  number={3},
  pages={034002--034002},
  year={2023},
  publisher={Society of Photo-Optical Instrumentation Engineers}
}

@article{si2024multiphonon,
  title={Multiphonon-assisted acousto-optical Q-switched laser at 1130 nm in Yb: YCOB crystal},
  author={Si, Huichen and Liang, Fei and Lu, Dazhi and Yu, Haohai and Zhang, Huaijin and Wu, Yicheng},
  journal={Chinese Optics Letters},
  volume={22},
  number={10},
  pages={101401},
  year={2024}
}

@article{jia2022integrated,
  title={Integrated photonics based on rare-earth ion-doped thin-film lithium niobate},
  author={Jia, Yuechen and Wu, Jiangwei and Sun, Xiaoli and Yan, Xiongshuo and Xie, Ranran and Wang, Lei and Chen, Yuping and Chen, Feng},
  journal={Laser \& Photonics Reviews},
  volume={16},
  number={9},
  pages={2200059},
  year={2022},
  publisher={Wiley Online Library}
}

@article{guo2021growth,
  title={Growth and optical properties of the Nd, Ce: YAG laser crystal},
  author={Guo, Yongwen and Huang, Jinqiang and Ke, Guanzhen and Ma, Yuanyuan and Quan, Jiliang and Yi, Guobin},
  journal={Journal of Luminescence},
  volume={236},
  pages={118134},
  year={2021},
  publisher={Elsevier}
}

@article{saleh2020improved,
  title={Improved Nd distribution in Czochralski grown YAG crystals by implementation of the accelerated crucible rotation technique},
  author={Saleh, Muad and Kakkireni, Saketh and McCloy, John and Lynn, Kelvin G},
  journal={Optical Materials Express},
  volume={10},
  number={2},
  pages={632--644},
  year={2020},
  publisher={Optical Society of America}
}

@article{de2022nd,
  title={Nd: YLF laser at 1053 nm diode side pumped at 863 nm with a near quantum-defect slope efficiency},
  author={de Almeida Vieira, T{\'a}rcio and Prado, Felipe Maia and Wetter, Niklaus Ursus},
  journal={Optics \& Laser Technology},
  volume={149},
  pages={107818},
  year={2022},
  publisher={Elsevier}
}

@article{turri2022mid,
  title={Mid-infrared spectroscopy of Nd: YLF crystal},
  author={Turri, G and Gennari, F and Bass, M and Toncelli, A},
  journal={Journal of Luminescence},
  volume={246},
  pages={118842},
  year={2022},
  publisher={Elsevier}
}

@article{zajic2023nd,
  title={Nd: YAG single crystals grown by the floating zone method in a laser furnace},
  author={Zajic, Frantisek and Klejch, Martin and Elias, Adam and Klicpera, Milan and Beitlerov{\'a}, Alena and Nikl, Martin and Pospisil, Jiri},
  journal={Crystal Growth \& Design},
  volume={23},
  number={4},
  pages={2609--2618},
  year={2023},
  publisher={ACS Publications}
}

@article{yu2025orthogonally,
  title={Orthogonally polarized dual-wavelength Nd: GdVO4/Nd: YVO4 laser at 1341 and 1342 nm with adjustable power ratio},
  author={Yu, Hao and Li, Yongliang and Moazzam, Fahad and Lin, Lin and Gao, Bo},
  journal={PloS one},
  volume={20},
  number={2},
  pages={e0317875},
  year={2025},
  publisher={Public Library of Science San Francisco, CA USA}
}

@article{song2021excellent,
  title={Excellent performance of a cryogenic Nd: YAlO3 laser with low wavefront distortion based on zero thermal expansion},
  author={Song, Yan-Jie and Xu, Yuan-Zhai and Meng, Shuai and Jiang, Xing-Xing and Shao, Chong-Feng and Song, Ze-Xin and Zong, Nan and Wang, Zhi-Min and Bo, Yong and Wang, Xiao-Jun and others},
  journal={Optics Letters},
  volume={46},
  number={10},
  pages={2425--2428},
  year={2021},
  publisher={Optical Society of America}
}

@article{li2023optically,
  title={Optically pumped milliwatt whispering-gallery microcavity laser},
  author={Li, Huiqi and Wang, Zhaocong and Wang, Lei and Tan, Yang and Chen, Feng},
  journal={Light: Science \& Applications},
  volume={12},
  number={1},
  pages={223},
  year={2023},
  publisher={Nature Publishing Group UK London}
}

@article{zhu2021integrated,
  title={Integrated photonics on thin-film lithium niobate},
  author={Zhu, Di and Shao, Linbo and Yu, Mengjie and Cheng, Rebecca and Desiatov, Boris and Xin, C\_J and Hu, Yaowen and Holzgrafe, Jeffrey and Ghosh, Soumya and Shams-Ansari, Amirhassan and others},
  journal={Advances in Optics and Photonics},
  volume={13},
  number={2},
  pages={242--352},
  year={2021},
  publisher={Optical Society of America}
}

@article{zhu2024twenty,
  title={Twenty-nine million intrinsic Q-factor monolithic microresonators on thin-film lithium niobate},
  author={Zhu, Xinrui and Hu, Yaowen and Lu, Shengyuan and Warner, Hana K and Li, Xudong and Song, Yunxiang and Magalh{\~a}es, Let{\'\i}cia and Shams-Ansari, Amirhassan and Cordaro, Andrea and Sinclair, Neil and others},
  journal={Photonics Research},
  volume={12},
  number={8},
  pages={A63--A68},
  year={2024}
}

@article{ruter2021optical,
  title={Optical characterization of a neodymium-doped lithium-niobate-on-insulator (LNOI)},
  author={R{\"u}ter, Christian E and Hasse, Kore and Chen, Feng and Kip, Detlef},
  journal={Optical Materials Express},
  volume={11},
  number={12},
  pages={4007--4014},
  year={2021},
  publisher={Optical Society of America}
}

@article{rosol2024short,
  title={Short pulsed (1088.5 nm) neodymium-doped fiber laser via Ti3C2TxPVA film-based SA},
  author={Rosol, AHA and Hamzah, A and Zulkipli, NF and Harun, SW},
  journal={Optik},
  volume={296},
  pages={171552},
  year={2024},
  publisher={Elsevier}
}

@article{najm2023generation,
  title={Generation of bright-dark pulses in a Q-switched thulium-doped fiber laser by using 8-HQCdCl2H2O},
  author={Najm, Mustafa Mohammed and Al-Hiti, Ahmed Shakir and Zhang, Pei and Al-Azzawi, Alabbas A and Nizamani, Bilal and Rosol, Ahmad Haziq Aiman and Abdullah, Mohammed Najm and Yasin, Moh and Harun, Sulaiman Wadi},
  journal={Optics \& laser technology},
  volume={164},
  pages={109450},
  year={2023},
  publisher={Elsevier}
}

@article{wang2025wgm,
  title={WGM Photonic Molecules as a Platform for Stable Pulsed Laser Emission Generation},
  author={Wang, Zhaocong and Cui, Qingqiang and Tan, Yang and Chen, Feng},
  journal={ACS Photonics},
  volume={12},
  number={10},
  pages={5528--5536},
  year={2025},
  publisher={ACS Publications}
}

@article{sun2017nonlinear,
  title={Nonlinear optical oscillation dynamics in high-Q lithium niobate microresonators},
  author={Sun, Xuan and Liang, Hanxiao and Luo, Rui and Jiang, Wei C and Zhang, Xi-Cheng and Lin, Qiang},
  journal={Optics express},
  volume={25},
  number={12},
  pages={13504--13516},
  year={2017},
  publisher={Optical Society of America}
}

@article{wu2025manipulation,
  title={Manipulation of Rare-Earth-Ion Emission by Nonlinear-Mode Oscillation in a Lithium Niobate Microcavity},
  author={Wu, Jiangwei and He, Yuxuan and Yang, Qilin and Wang, Xueyi and Liu, Xiangmin and Geng, Yong and Guo, Guangcan and Zhou, Qiang and Chen, Xianfeng and Chen, Yuping},
  journal={Nano Letters},
  year={2025},
  publisher={ACS Publications}
}

@article{song2025stable,
  title={Stable gigahertz-and mmWave-repetition-rate soliton microcombs on X-cut lithium niobate},
  author={Song, Yunxiang and Zhu, Xinrui and Zuo, Xiangying and Huang, Guanhao and Lon{\v{c}}ar, Marko},
  journal={Optica},
  volume={12},
  number={5},
  pages={693--701},
  year={2025},
  publisher={Optica Publishing Group}
}

@article{nie2025soliton,
  title={Soliton microcombs in X-cut LiNbO3 microresonators},
  author={Nie, Binbin and Lv, Xiaomin and Yang, Chen and Ma, Rui and Zhu, Kaixuan and Wang, Ze and Liu, Yanwu and Xie, Zhenyu and Jin, Xing and Zhang, Guanyu and others},
  journal={eLight},
  volume={5},
  number={1},
  pages={15},
  year={2025},
  publisher={Springer}
}

@article{he2023massively,
  title={Massively parallel FMCW lidar with cm range resolution using an electro-optic frequency comb},
  author={He, Bibo and Zhang, Chenbo and Yang, Jiachuan and Chen, Nuo and He, Xuanjian and Tao, Jinming and Zhang, Zhike and Chu, Tao and Chen, Zhangyuan and Xie, Xiaopeng},
  journal={Optics Letters},
  volume={48},
  number={13},
  pages={3621--3624},
  year={2023},
  publisher={Optica Publishing Group}
}

@article{wang2023nanosecond,
  title={Nanosecond-pulsed passively q-switched fiber laser by using photothermal dynamics in a dielectric microcavity},
  author={Wang, Wenyu and Xiao, Bowen and Zhao, Ping and Ren, Linhao and Zhu, Song and Shi, Lei and Zhang, Xinliang},
  journal={ACS Photonics},
  volume={10},
  number={10},
  pages={3656--3663},
  year={2023},
  publisher={ACS Publications}
}

@article{fu2024soliton,
  title={Soliton microcomb generation by cavity polygon modes},
  author={Fu, Botao and Gao, Renhong and Yao, Ni and Zhang, Haisu and Li, Chuntao and Lin, Jintian and Wang, Min and Qiao, Lingling and Cheng, Ya},
  journal={Opto-Electronic Advances},
  volume={7},
  number={8},
  pages={240061--1},
  year={2024}
}

@article{wu2018lithium,
  title={Lithium niobate micro-disk resonators of quality factors above 107},
  author={Wu, Rongbo and Zhang, Jianhao and Yao, Ni and Fang, Wei and Qiao, Lingling and Chai, Zhifang and Lin, Jintian and Cheng, Ya},
  journal={Optics letters},
  volume={43},
  number={17},
  pages={4116--4119},
  year={2018},
  publisher={Optical Society of America}
}

@article{liu2021chip,
  title={On-chip erbium-doped lithium niobate microcavity laser},
  author={Liu, YiAn and Yan, XiongShuo and Wu, JiangWei and Zhu, Bing and Chen, YuPing and Chen, XianFeng},
  journal={Science China Physics, Mechanics \& Astronomy},
  volume={64},
  number={3},
  pages={234262},
  year={2021},
  publisher={Springer}
}

@article{liu2025ultralow,
  title={Ultralow-Threshold Lithium Niobate Photonic Crystal Nanocavity Laser},
  author={Liu, Xiangmin and Chen, Chengyu and Ge, Rui and Wu, Jiangwei and Chen, Xianfeng and Chen, Yuping},
  journal={Nano Letters},
  volume={25},
  number={16},
  pages={6454--6460},
  year={2025},
  publisher={ACS Publications}
}

@article{zhang2021integrated,
  title={Integrated lithium niobate single-mode lasers by the Vernier effect},
  author={Zhang, Ru and Yang, Chen and Hao, ZhenZhong and Jia, Di and Luo, Qiang and Zheng, DaHuai and Liu, HongDe and Yu, XuanYi and Gao, Feng and Bo, Fang and others},
  journal={Science China Physics, Mechanics \& Astronomy},
  volume={64},
  number={9},
  pages={294216},
  year={2021},
  publisher={Springer}
}

@article{chen2022photonic,
  title={Photonic integration on rare earth ion-doped thin-film lithium niobate},
  author={Chen, Yuping},
  journal={Science China. Physics, Mechanics \& Astronomy},
  volume={65},
  number={9},
  pages={294231},
  year={2022},
  publisher={Springer Nature BV}
}

\end{document}